\documentclass[twocolumn,prl,preprintnumbers,amsmath,amssymb,groupedaddresst]{revtex4}
\usepackage{graphicx}

\usepackage{color}
\usepackage{bm}
\begin{document}

\preprint{AP-GR-114, OCU-PHYS-409, OU-HET-821, RIKEN-MP-93}

\title{Turbulent meson condensation in quark deconfinement}

\author{Koji Hashimoto$^{1,2}$}
\author{Shunichiro Kinoshita$^{3}$}
\author{Keiju Murata$^{4}$}
\author{Takashi Oka$^{5}$}
\affiliation{$^{1}${\it Department of Physics, Osaka University,
Toyonaka, Osaka 560-0043, Japan}}
\affiliation{$^{2}${\it Mathematical Physics Lab., RIKEN Nishina Center,
Saitama 351-0198, Japan}}
\affiliation{$^{3}${\it Osaka City University Advanced Mathematical Institute, Osaka 558-8585, Japan}}
\affiliation{$^{4}${\it Keio University, 4-1-1 Hiyoshi, Yokohama 223-8521, Japan}}
\affiliation{$^{5}${\it Department of Applied Physics, University of Tokyo, 
Tokyo 113-8656, Japan}}

\begin{abstract}
In a QCD-like strongly coupled gauge theory at large $N_c$, using the AdS/CFT correspondence, 
we find that 
heavy quark deconfinement is accompanied by a coherent condensation of higher meson resonances.
This is revealed in non-equilibrium deconfinement transitions triggered by 
static, as well as, quenched electric fields even below the Schwinger limit. 
There, we observe a ``turbulent'' energy flow 
to higher meson modes,
which finally results in the quark deconfinement.
Our observation is consistent with seeing deconfinement as a condensation of long QCD strings.
\end{abstract}

\maketitle



Quark confinement is one of the most fundamental and challenging problems in elementary particle physics, left unsolved. Although quantum chromodynamics (QCD) 
is the fundamental field theory describing quarks and gluons, 
their clear understanding is limited to the deconfined phase
at high energy or high temperature limits due to the asymptotic freedom. 
We may benefit from employing a more natural description of the zero temperature hadron vacuum.
A dual viewpoint of quark confinement in terms of the ``fundamental'' 
degrees of freedom at zero temperature - mesons, is a plausible option. 

The mesons appear in families: they are categorized by their spin/flavor quantum numbers, 
as well as a resonant excitation level $n$ giving a resonance tower such as $\rho(770), \rho(1450),\rho(1700), \rho(1900), \cdots$.
In this Letter we find a novel behavior of the higher meson resonances, {\it i.e.}, mesons with large $n$.
In the confined phase,
when the deconfined phase is approached, we observe {\it condensation of higher mesons}. 
In this state, macroscopic number of the higher meson resonances, with a characteristic distribution, are excited.
The condensed mesons have the same quantum number as the vacuum. 
The analysis is done via the anti-de Sitter space (AdS)/conformal field
theory (CFT) correspondence \cite{Maldacena:1997re,Gubser:1998bc,Witten:1998qj},
one of the most reliable tools to study strongly-coupled gauge theories.
By shifting our viewpoint from quark-gluon to meson degrees of freedom, we gain a simple 
and universal understanding of the confinement/deconfinement transition, 
with a bonus of solving mysteries in black holes physics through the AdS/CFT.

The system we study is the ${\cal N}=2$ supersymmetric $SU(N_c)$ QCD
which allows the simplest AdS/CFT treatment \cite{Karch}.
The deconfinement transition is induced by external electric fields
\footnote{Introducing temperature may generate the same effect, which we
will report soon \cite{Next}.}. In static fields, the confined phase 
becomes unstable in electric fields stronger than the 
Schwinger limit $E=E_\textrm{Sch}$ beyond which quarks are liberated from the confining force. 
We find that this instability is accompanied by the condensation of higher mesons.  
A striking feature is revealed  for the case of an electric field quench:
The kick from the quench triggers a domino-like energy transfer from low to high
resonant meson modes.
This leads to a dynamical deconfinement transition \cite{Hashimoto:2014yza}
even below the Schwinger limit. 
The transfer we find resembles that of turbulence in classical hydrodynamics 
as higher modes participate;
thus we call it a ``turbulent meson condensation" and suggest it being 
responsible for deconfinement. 

We remind that the ${\cal N}=2$ theory is a toy model:  The meson sector is confined and 
has a discrete spectrum while the gluon sector is conformal and is always deconfined.   
It resembles heavy quarkonia in a gluon plasma. 
Generically, quark deconfinement and gluon deconfinement can happen separately, as is known through charmonium experiments in heavy ion collisions. Here, we concentrate on the deconfinement of
heavy quarks and not the gluons.

The higher meson resonances are naturally interpreted as long QCD strings, therefore
our finding is consistent with interpreting deconfinement as
condensation of QCD strings \cite{Polyakov:1978vu} 
(see \cite{Lucini:2005vg,Hanada:2014noa,Pisarski:1982cn,Patel:1983sc}). 
Under the condensation, a quark can propagate away from its partner antiquark by 
reconnecting the bond QCD string with the background condensed strings.
The gravity dual of the deconfined phase is with a  black hole, so
given the relation with long fundamental strings \cite{Hanada:2014noa}, our result 
may shed light on the issue of quantum black holes; In particular, 
our time-dependent analysis gives a singularity formation on the flavor D-brane in AdS,
a probe-brane version of the Bizon-Rostworowski turbulent instability in
AdS geometries \cite{Bizon:2011gg}. 


\vspace{3mm}
\noindent
{\it Review: Meson effective action from AdS/CFT.} ---
The effective field theory of mesons can be obtained 
for the ${\cal N}=2$ supersymmetric QCD
in the large $N_c$,  $\lambda \equiv N_c g_{\rm YM}^2$ limits 
by the AdS/CFT correspondence\cite{Kruczenski:2003be,Erdmenger:2007cm}.
The meson action is nothing but a D7-brane action in the 
$\mathrm{AdS}_5\times S^5$ geometry;
\begin{eqnarray}
&&
S =  \frac{-1}{(2\pi)^6 g_{\rm YM}^2 \l_s^8}\int \!\! d^8\xi \sqrt{
-\det (g_{ab} [w]+ 2\pi l_s^2 F_{ab})} \, ,
\label{D7action}
\\
&&
ds^2 = \frac{r^2}{R^2}\eta_{\mu\nu}dx^\mu dx^\nu \!+\! \frac{R^2}{r^2}
\!\left[
d\rho^2 \!+\! \rho^2 d\Omega_3^2 \!+\! dw^2\!+\!d\bar{w}^2
\right]\, ,
\nonumber
\end{eqnarray}
where $r^2 \equiv \rho^2 + w^2 + \bar{w}^2$, 
$F_{ab}=\partial_a A_b-\partial_b A_a$, 
and the $\mathrm{AdS}_5$ curvature radius is 
$R\equiv (2\lambda)^{1/4} l_s$.
For the following calculations, it is convenient to define a rescaled gauge potential 
$a_a\equiv 2\pi l_s^2 R^{-2} A_a$.
The D7-brane worldvolume fields are
$w(x^\mu,\rho)$ and $a_a(x^\mu,\rho)$.
(We set $\bar{w}(x^\mu,\rho)=0$ consistently because of
$U(1)$-symmetry in ($w,\bar{w}$)-plane.)
We denote the location of
the D7-brane at the asymptotic AdS boundary
as $w(x^\mu,\rho=\infty)=R^2 m$.
Here, the constant $m$ is related to the quark mass $m_q$ as
$m_q=(\lambda/2\pi^2)^{1/2}m$.
A static solution of the D7-brane in the 
$\mathrm{AdS}_5\times S^5$ geometry is given by 
$w(x^\mu,\rho)=R^2m$ and $a_a(x^\mu,\rho)=0$. 
Using the AdS/CFT dictionary, normalizable fluctuations around the static solutions 
of $w$ and $a_a$
are interpreted as the infinite towers
of scalar mesons $\bar\psi \psi$ and vector mesons
$\bar\psi \gamma_\mu \psi$, respectively. 
(We omit the pseudo scalar mesons which
are irrelevant to our discussion).
In this paper, we focus on the meson condensation induced by an electric
field along the $x$-direction. 
Thus, we only consider fluctuations of $w$ and $a_x$.

To derive the meson effective action, 
we use a coordinate $z$ defined by
$\rho= R^2m\sqrt{1-z^2}/z$
(where
$z=0$ is the AdS boundary, and $z=1$ is the D7-brane center that is closest to
the Poincar\'e horizon in the bulk AdS).
The worldvolume is effectively in a finite box along the AdS radial direction,
to give a confined discrete spectrum as we will see below.
We expand the D7-brane action 
up to second order in the fluctuations
$\bm{\chi}\equiv (R^{-2}w-m,a_x)$ as \cite{Kruczenski:2003be}
\begin{equation}
S= \int \!dt d^3x\int^1_0 \!dz \frac{1-z^2}{2z}[
\dot{\bm{\chi}}^2-m^2(1-z^2)\bm{\chi}'{}^2
]\ + {\cal O} (\bm{\chi}^3),
\nonumber
\end{equation}
where ${}^\cdot\equiv \partial_t$ and $'\equiv \partial_z$.
An irrelevant overall factor is neglected.
The equation of motion for $\bm{\chi}$ is
\begin{equation}
 \left(\partial_t^2+\mathcal{H}\right)\bm{\chi}=0\ ,
 \quad
 \mathcal{H}\equiv-
 m^2
 \frac{z}{1\!-\!z^2}
 \frac{\partial}{\partial z}\frac{(1\!-\!z^2)^2}{z}
 \frac{\partial}{\partial z}
 \ .
\label{lineq}
\end{equation}
The eigenfunction of $\mathcal{H}$ is given by
$e_n(z)\equiv \sqrt{2(2n+3)(n+1)(n+2)} \, z^2 F(n+3,-n,2;z^2)$,
with the eigenvalue
$\omega_n^2=4(n+1)(n+2)m^2$,
for the meson level number $n=0,1,2,\cdots$.
Here $F$ is the Gaussian hypergeometric function. 
The  inner product is defined as
$(f,g)\equiv \int^1_0 dz\, z^{-1}(1-z^2)f(z)g(z)$, 
where 
 $(e_n,e_m)=\delta_{mn}$ is satisfied.
Note that an external electric field term, $a_x=-Et$, 
satisfies Eq.~(\ref{lineq}) although it is non-normalizable.
Expanding the scalar/vector fields as 
\begin{eqnarray}
\bm{\chi}=(0,-Et)+\sum_{n=0}^\infty \bm{c}_n(t)e_n(z)\, , 
\label{decompomeson}
\end{eqnarray}
we find an infinite tower of meson fields $\bm{c}_n(t)$ sharing the same quantum charge -
higher meson resonances.
Substituting
Eq.~(\ref{decompomeson})
back to Eq.~(\ref{D7action}), we obtain the meson effective action
\begin{eqnarray}
S = \frac12 \int \! d^4x\;
\sum_{n=0}^\infty \left[\dot{\bm{c}}_n^2 - \omega_n^2 \bm{c}_n^2 \right] + \mbox{interaction} \, ,
\label{mesoneff}
\end{eqnarray}
where we have omitted a constant term and total derivative terms,
while higher order nonlinear terms give rise to meson-meson interactions. From 
the effective action, 
we obtain the energy $\varepsilon_n\equiv \frac{1}{2}(\dot{\bm{c}}_n^2+\omega_n^2 \bm{c}_n^2)$ stored in the $n$-th meson
resonance, 
and the linearized total energy $\varepsilon=\sum_{n=0}^{\infty} \varepsilon_n$.


\vspace{3mm}
\noindent
{\it Higher meson condensation and deconfinement.}---
The confined phase becomes unstable in strong static electric fields. 
Here, we examine this from the viewpoint of meson condensation. 
In Eq.~(\ref{mesoneff}), the meson couplings depend
on the external electric field $E$ nonlinearly. 
Mesons in a single flavor theory is neutral under $E$, but it can polarize,
and non-linear $E$ may cause a meson condensation.
We first solve the equations of motion
obtained from the full nonlinear 
D-brane action (\ref{D7action}) with an external electric field, 
and then decompose the solution
$\bm{\chi}(t,z)$ as Eq.~(\ref{decompomeson}). 
In this way, we can study how the infinite tower of mesons
 $\bm{c}_n(t)$ behave towards the deconfinement transition.

\begin{figure}
\includegraphics[width=6.5cm]{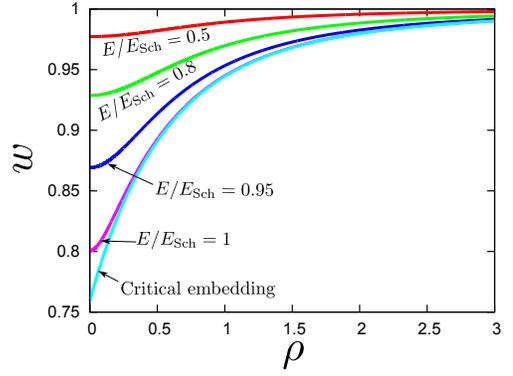}
\caption{(Color online) The shape of the probe D7-brane in static electric fields
in the unit of $R=m=1$. 
The lines correspond respectively to
$E/E_\textrm{Sch}=0.5, 0.8, 0.95, 1$ and the critical embedding from top to bottom.
}
\label{figEshape}
\end{figure}
\begin{figure}
\includegraphics[width=6.7cm]{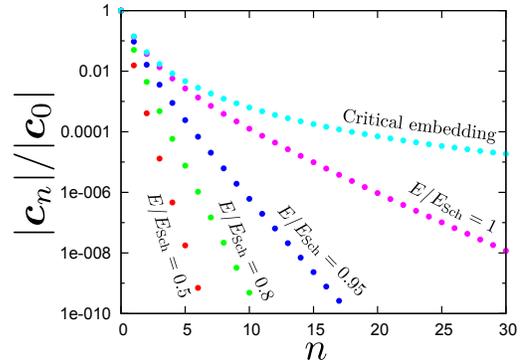}
\caption{(Color online) Decomposed meson condensate 
$\log[|\bm{c}_n|/|\bm{c}_0|]$ in static electric fields. 
Colors Red/Green/Blue/Magenta/Cyan 
correspond respectively to $E/E_\textrm{Sch}=0.5, 0.8, 0.95, 1$
and the critical embedding.
}
\label{figEdecomp}
\end{figure}

The static D7-brane solution 
in the presence of a constant electric field 
introduced by $a_x=-Et$ was 
obtained in Refs.~\cite{Karch:2007pd,Albash:2007bq,Erdmenger:2007bn}. 
Also, the Schwinger limit $E=E_\textrm{Sch}=0.5759m^2$ beyond
which the first order phase transition to deconfinement occurs was
found~\cite{Albash:2007bq,Erdmenger:2007bn}.
Fig.~\ref{figEshape} shows the shape of the D7-brane, which is the
scalar field configuration $w(\rho)$, for
$E/E_\textrm{Sch}=0.5,0.8,0.95,1$ and the critical embedding.
At the critical embedding which is a confinement/deconfinement 
phase boundary although the solution itself is 
thermodynamically unfavored, the brane becomes conical and an effective
horizon starts emerging on the worldvolume.
As the solution approaches the critical embedding, 
the D7-brane bends more toward the Poincar\'e horizon of AdS;
the brane with the sharpest bending is for the critical embedding.

The bending leads to meson condensation, where
sharp bending means that higher meson modes are excited. 
In order to clarify this, 
we decompose the scalar field by the meson eigenmodes and 
plot the ratio $|\bm{c}_n|/|\bm{c}_0|$ as a function of $n$ 
in Fig.~\ref{figEdecomp},  where we define 
$|\bm{c}_n| \equiv \sqrt{(c_n^{\rm scalar})^2+(c_n^{\rm vector})^2}$ for illustration.
As the electric field increases, the higher meson condensate $|\bm{c}_n|$ 
$(n\gg 1)$ compared to the lowest $|\bm{c}_0|$ grows rapidly. 
Note that vector mesons are not excited in static electric fields 
since the gauge potential is always given by $a_x=-Et$ (no higher modes). 
This implies that 
the condensed mesons have the same quantum number as the vacuum.
The geometrical reason of the condensation of the higher mesons is
simple; The D7-brane bends singularly near the 
Poincar\'e horizon of AdS due to the nature of the metric
and higher eigenmodes are necessary to reproduce it. 
It is similar to Fourier-decompose a delta-function (narrow Gaussian).

\begin{figure} 
\includegraphics[scale=0.45]{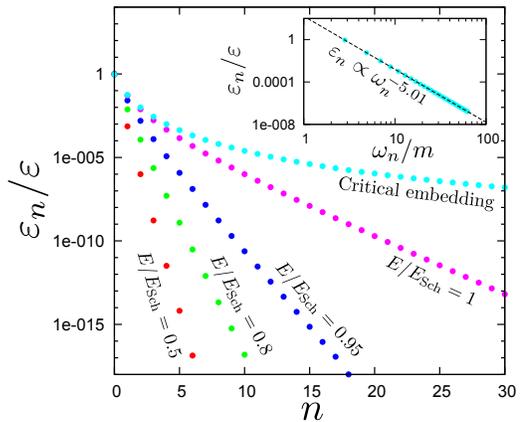}
\caption{(Color online) The energy distribution for the $n$-th meson resonance.
The color of the dots follows that of the previous figure.
The inset is the log-log plot of the energy distribution for the
 critical embedding, in which 
we take the meson mass spectrum $\omega_n$ as the horizontal axis.
}
\label{finiteEenergy}
\end{figure}

``Flow'' of energy from low to high resonant meson modes takes place at the same time.
For static solutions, the energy stored by the $n$-the meson
mode is given by $\varepsilon_n=\omega_n^2 \bm{c}_n^2/2$.
In Fig.~\ref{finiteEenergy}, 
the meson energy distribution in static $E$
shows enhancement at higher
modes as increasing $E$, that is, energy ``flow'' to higher meson resonances.
The exponential behavior of the enegy distribution in Fig.~\ref{finiteEenergy} for $E_{\rm Sch}\geq E$ provides a well-defined effective temperature \cite{Next}. At 
the critical embedding it turns to a power-law, which exhibits a 
Hagedorn behavior, and reminds us of the Kolmogorov scaling.

We conclude for the static case
that just before the quark deconfinement induced by an applied electric field,
meson resonances condense coherently.

\vspace{3mm}
\noindent
{\it Meson turbulence in quenched electric fields.}---
The higher meson condensation seems to be a 
sufficient cause of quark deconfinement.
This is clearly seen in a time-dependent, electric field quench
that we study below. 
Starting from the $E=0$ vacuum in the confined phase,
we turn on the electric field smoothly to reach a final value $E_f$ in the duration 
$\Delta V$ \footnote{The profile is  
$E = E_f [V-\frac{\Delta V}{2\pi} \sin(2\pi V/\Delta V)]/\Delta V$ \cite{Hashimoto:2014yza}.}.
In our previous work \cite{Hashimoto:2014yza}, we 
found that the system deconfines with an emergence 
of a strongly redshifted region
on the D7-brane toward a naked-singularity formation.
This is interpreted as an instability toward deconfinement, 
which happens, to our surprise, even when the final 
field strength is below the Schwinger limit.
In the following, we choose a weak electric field 
($E_f/E_\textrm{Sch}=0.2672$) and a switch-on duration 
$m \Delta V=2$, in which 
sub-Schwinger-limit deconfinement is realized.

\begin{figure}
  \centering
  \includegraphics[scale=0.45]{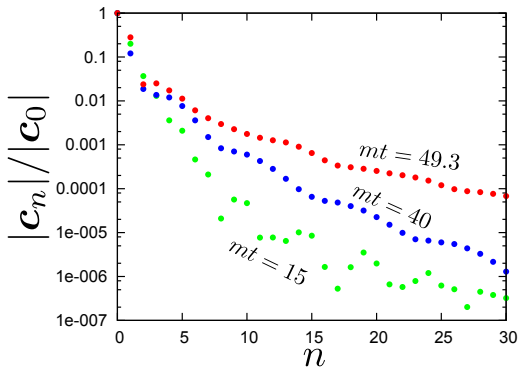} 
  \caption{
(Color online) Meson condensate induced by an electric field quench. 
Data for times $m t=15$, $40$, and $49.3$ are shown.
We set 
$E_f/E_\textrm{Sch}=0.2672$ 
and $m \Delta V=2$.
A clear (non-thermal) growth at large $n$ is found along the time evolution.  
}
\label{nth_cond}
\end{figure}

\begin{figure}
  \centering
  \includegraphics[scale=0.45]{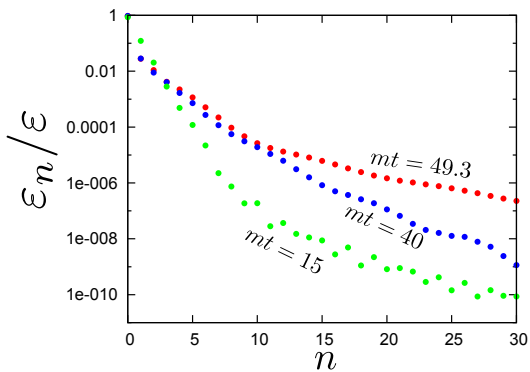}
  \caption{
(Color online) Turbulent behavior of mesons toward deconfinement.
The energy is
transferred to higher modes accordingly.
}
\label{nth_energy}
\end{figure}

The electric field induces nontrivial 
dynamics in the meson sector.
We decompose the time-dependent solutions of the
meson fields~\cite{Hashimoto:2014yza}
and calculate the energy
$\varepsilon_n$ 
of the meson resonances.
In Fig.~\ref{nth_cond}, the time evolution of the 
condensate $|\bm{c}_n|$ is shown
for several time slices $mt=15$, $40$, and $49.3$, while
the time $m t=49.3$ is just before deconfinement.
Figure~\ref{nth_energy} plots
$\varepsilon_n/\varepsilon$. We find that the condensate and the energy  
are transferred to 
higher meson modes during the time evolution.
This tendency is similar to the static case
in which they are ``transferred'' more in stronger electric
fields. This indicates that 
higher meson condensation is universally 
related to quark confinement. 

The observed time evolution of the distribution suggests that turbulence is taking place in the meson sector. 
This is because higher modes have smaller wave lengths, 
and the transfer of energy and momentum indicates that smaller structure are being organized 
during the time evolution toward deconfinement. 
Our finding can be considered as a probe-brane version of the turbulent instability of the AdS 
spacetime \cite{Bizon:2011gg} in which a non-linear evolution of 
a perturbed AdS spacetime causes a high-momentum instability resulting in a black hole formation.


\vspace{3mm}
\noindent
{\it Conclusion and discussions.}---
In this work we found, via the AdS/CFT correspondence, 
that higher meson resonances become condensed
near the deconfinement transition caused by electric fields. 
This was confirmed for
both the static electric fields and the time-dependent quenched electric field,
in ${\cal N}=2$ supersymmetric QCD which models
heavy quarkonia in a gluon plasma.
No internal symmetry is broken during this process 
since 
the mesons that participates in the deconfinement with their condensation
have the same quantum numbers as the vacuum and the external force.

The physics in the gravity dual side is simple. 
The gravity-dual of the scalar meson condensation is the deformation of the 
probe D7-brane. The electric field breaks the supersymmetries of the brane configuration
making the Poincar\'e horizon to attract the D7-brane and to bend it.
The tip of the D7-brane becomes sharper as the electric field increases,
and finally when it exceeds the critical value, deconfinement transition occurs.
At the verge of deconfinement (at which we can use the meson terminology),
higher meson resonances condense coherently.

A single meson can be regarded as quarks connected by a QCD string
and higher meson resonances as its coherent fluctuation. 
At the deconfinement transition,
one needs to dissociate the QCD string to liberate the quarks. 
The dissociation of the QCD string is described by the coherent dynamics of the excited mesons. 
Our finding is consistent with this view,
and in particular with deconfinement as QCD string condensation
\cite{Hanada:2014noa, Lucini:2005vg,Polyakov:1978vu,Pisarski:1982cn,Patel:1983sc}. 
Our result seems further consistent with Hagedorn transition in string theory 
\cite{Atick:1988si, Hagedorn:1965st} \footnote{K.H.~thanks S.~Yamaguchi 
for pointing out its relevance to Hagedorn transition.} and
the black hole / string correspondence \cite{Susskind:1993ws, Horowitz:1996nw}.

In quark models, the higher resonant meson naively corresponds to a state 
$\bar{\psi}(x) U(x,y)\psi(y)$, 
where $U(x,y)\equiv {\rm P} \exp[i\int_x^y G]$  is an open Wilson line
with the gluon field $G_\mu$,
which can be expanded as $\bar\psi(x) \Box^n \psi(x)$ where $\Box$ is the covariant Laplacian.
Our conjecture waits for a confirmation by lattice QCD on condensation of this operator.

The meson turbulence obviously share some features with the AdS turbulence
\cite{Bizon:2011gg,Dias:2011ss}.
It may have relation with thermalization and numerical simulations of
glasma \cite{McLerran:1993ni,McLerran:1993ka,McLerran:1994vd}. 
Our observation is also 
valid in a temperature-driven deconfinement \cite{Next}.
Thus, we conjecture that quark deconfinement is universally associated with 
a condensation of higher meson resonances.


\vspace*{2mm}
{\noindent \it Acknowledgment.} ---
K.~H.~would like to thank A.~Buchel, H.~Fukaya, D.-K.~Hong, N.~Iqbal, K.-Y.~Kim, J.~Maldacena, S.~Sugimoto, S.~Yamaguchi and P.~Yi for valuable discussions, and APCTP focus week program for its hospitality.  
This research was
partially supported by the RIKEN iTHES project.


\end{document}